# New applications of Equinox code for real-time plasma equilibrium and profile reconstruction for tokamaks

K. Bosak [a], J. Blum [a], E. Joffrin [b]

[a] *Laboratoire J. A. Dieudonné, Université de Nice-Sophia-Antipolis, 06100 Nice, France*
[b] *Association EURATOM/CEA sur la fusion, CEA Cadarache, F13108 St Paul-lez-Durance, France*

ABSTRACT

Recent development of real-time equilibrium code Equinox [1] using a fixed-point algorithm [2] allow major plasma magnetic parameters to be identified in real-time, using rigorous analytical method [3]. The code relies on the boundary flux code providing magnetic flux values on the first wall of vacuum vessel. By means of least-square minimization of differences between magnetic field obtained from previous solution and the next measurements the code identifies the source term of the non-linear Grad-Shafranov equation [4]. The strict use of analytical equations together with a flexible algorithm offers an opportunity to include new measurements into stable magnetic equilibrium code and compare the results directly between several tokamaks while maintaining the same physical model (i.e. no iron model is necessary inside the equilibrium code).

The successful implementation of this equilibrium code for JET and Tore Supra has already been published [1]. In this paper, we show the preliminary results of predictive runs of the Equinox code using the ITER geometry.

Because the real-time control experiments of plasma profile at JET using the code has been shown unstable when using magnetic and polarimetric measurements (that could be indirectly translated into accuracy vs robustness tradeoff), we plan an outline of the algorithm that will allow us to further constrain the plasma current profile using the central value of pressure of the plasma in real-time in order to better define the poloidal beta (this constraint is not necessary with purely magnetic equilibrium).

MOTIVATION

The goal of a real-time equilibrium code is to identify:
- the plasma boundary
- the flux surface geometry outside and inside of the plasma
- the current density profile
- derive safety factor profile and other important parameters from obtained equilibrium

In order to meet the real-time requirements, since 1999 an entirely new code EQUINOX has been designed and implemented in C++ using the latest software engineering techniques. Its application to various tokamaks and integration with existing codes follows.

MATHEMATICAL LAYOUT AND ALGORITHM

In order to find the plasma equilibrium, we solve the following stationary, nonlinear, bidimensional differential Grad-Shafranov equation:

$$-\Delta^* \psi(r,z) = J_P(r, \overline{\psi}(r,z)) \tag{1}$$

$$\Delta^* = \frac{\partial}{\partial r}\left[\frac{1}{\mu_0 r}\frac{\partial}{\partial r}\right] + \frac{\partial}{\partial z}\left[\frac{1}{\mu_0 r}\frac{\partial}{\partial z}\right]$$

where $\overline{\psi}$ is the normalized poloidal magnetic flux

The right-hand side of the equation is composed of the two functions to be identified:
Pressure function P and diamagnetic function F.

$$J_P(r,\psi) = rP'(\psi) + \frac{1}{\mu_0 r} F(\psi) F'(\psi) \tag{2}$$

The use of Picard type (fixed-point) algorithm allows the calculation of the inverse of finite element stiffness matrix, while the limitation to the first wall reduces mesh size and removes the nonlinearity of the magnetic permeability of iron at JET. Global optimisation of the code during its creation allows further acceleration, finally obtaining a code that is 2-3 orders of magnitude faster than offline equilibrium codes.

The following formulation (3) of the equation (1) forms the basis of finite element method:

Find $\psi \in H^1(\Omega)$ such that

$$\int_\Omega \frac{1}{\mu_0 r} \nabla \psi \nabla v \, d\Omega = \lambda \int_{\Omega_P} j_P(r, \overline{\psi}) v \, d\Omega \quad \forall v \in H_0^1(\Omega) \tag{3}$$

It can be written as :

$$K \underline{\psi}^{n+1} = B(\psi^n, \Omega_P^{\ n}) \underline{u}^n - \underline{h} \tag{4}$$

where: K – stifness matrix, B – source term matrix, u – vector of fluxes at nodes of the mesh, h – a vector modifying source term because of Dirichlet boundary conditions (the values of poloidal fluxes), n – iteration index

The fitting problem can be written as:

$$C \underline{\psi}^{n+1} = C(K^{-1} B(\psi^n, \Omega_P^{\ n}) \underline{u}^n - K^{-1} \underline{h}) \cong \underline{g} \tag{5}$$

where: g – Neumann boundary conditiotions (poloidal magnetic field, plus ev. faraday rotation angles or MSE measurements), C – magnetic (+ev. polarimetry) measurement matrix
Please note that if the fluxes are known from previous iteration, the left-hand sidematrix can be precomputed:

$$CK^{-1} B(\psi^n, \Omega_P^{\ n}) \underline{u}^n \cong \underline{g} + CK^{-1} \underline{h} \tag{6}$$

This leads to very efficient algorithm. In the real-time versions of the code a perturbation algorithm is used. The perturbation algorithm uses direct solver to calculate the residuals for a few cases close to the previous solution. The minimisation of the cost function is performed using binomial approximation of the minimum bracketed by neighbouring solutions. This prevents the code from choosing globally optimal but transient and potentially divergent solutions.

The code relies on the boundary code reconstruction providing total plasma current, toroidal magnetic field, magnetic flux values and poloidal magnetic field on the first wall of vacuum vessel. This improves the portability of the code, since we are not asking for the boundary itself.
From the practical point of view, the boundary codes are traditionally required to give an accurate plasma boundary even if this boundary sometimes is discontinuous or we can observe local oscillations due to high-degree polynomial extrapolation used in those codes. Hopefully, while using the boundary conditions on the first wall we use the values where they are the most exact, while hiding tokamak magnetic measurement-specific issues in boundary codes. Fig. 1. illustrates this approach for JET and ToreSupra tokamaks.

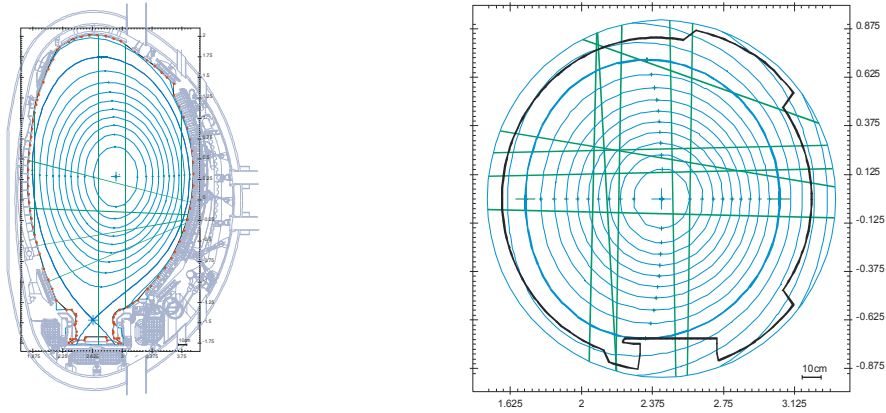

Fig. 1.

More specifically the solution provided by Equinox encompasses the scrape-off layer, while supposing the plasma current density to be null in this region. Because the plasma boundary found under those assumptions is close to the results from widely accepted boundary control codes like XLOC or BETALI (+/- 1cm at JET at the worst case) we found this assumption to be numerically justified. On the other hand the electronic density is supposed constant and non-vanishing outside of the plasma – this has been justified by the direct measurement of LIDAR at JET: even when the electronic density profile was consistent (+/-10%, no bias) vs. LIDAR, the global volume average was too low for Equinox results when compared with the value measured by the LIDAR system. After introducing the electronic density in the scrape-off layer the bias has been removed. The model of keeping constant electronic density may seem to be quite primitive but no other direct measurements are indicating the need of improvement, and remaining electronic density fitting residuals are now at the level of measurement noise.

TYPES OF MEASUREMENTS IN USE

There are three major versions of the code:
1. EQUINOX-D, very fast direct Shafranov solver
2. EQUINOX-M, using magnetic measurements only, gives accurate plasma geometry and optionally electronic density (with interferometry) and other profiles (Ti, Te...) .
3. EQUINOX-J using magnetics and polarimetry in order to identify and control hollow plasma profiles. In order to validate the code, the whole chain of JET codes has been explored (Fig. 2.)

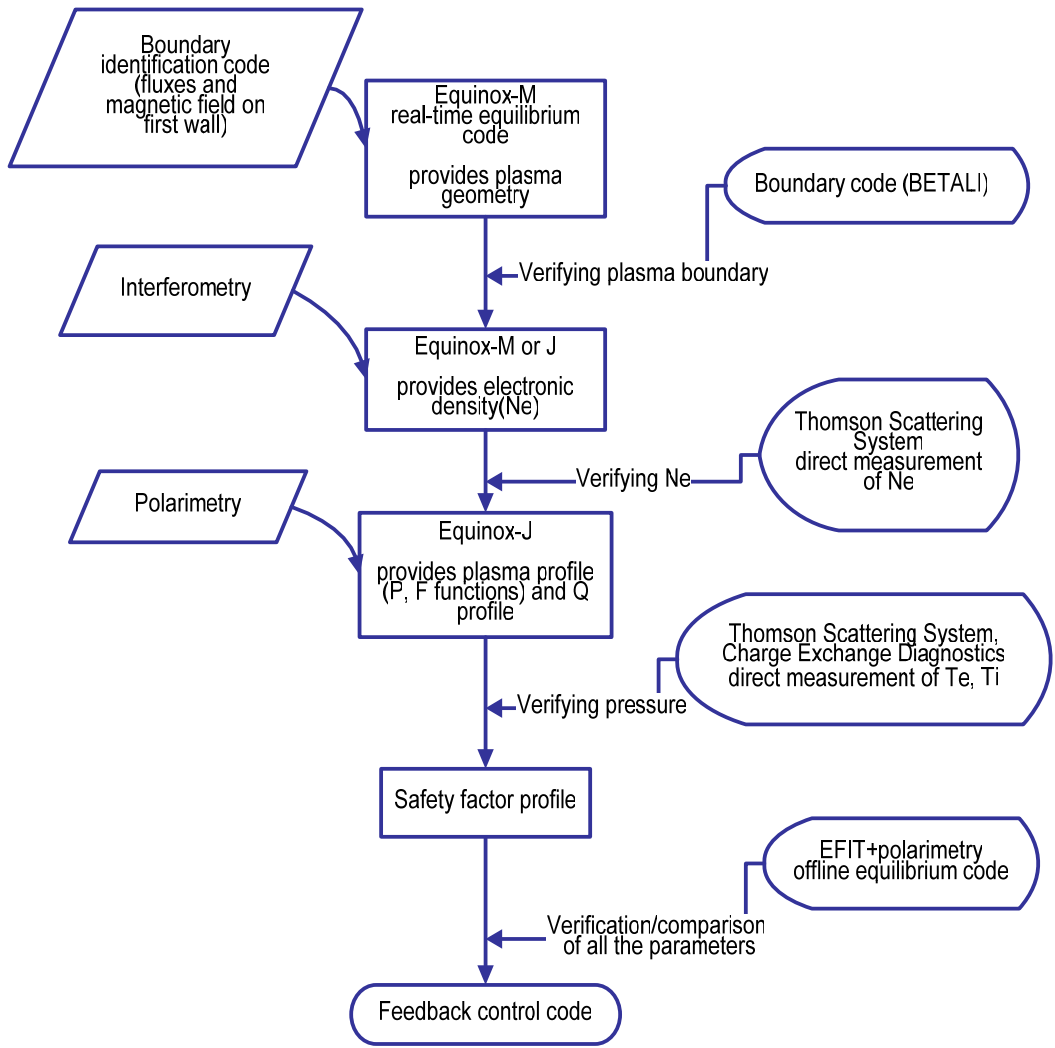

Fig. 2.

EQUINOX-D is currently being introduced into CRONOS code as a replacement of equilibrium code HELENA.

NEW APPLICATIONS - ITER RESULTS

EQUINOX has been tested on ITER equilibria. Since the pure boundary code for ITER does not exist, there are 2 possible use scenarios:
1. In predictive mode, after furnishing a mesh encompassing desired plasma geometry, the code can be used as a direct Shafranov solver or as a very precise plasma current profile identification code (magnetic measurements are on the plasma boundary in this scenario) (Fig. 3.)
2. Using the same approach as with the other tokamaks, the code can generate the equilibria using the magnetic measurements on the first wall. This geometry would allow the use of interferometry and polarimetry. In predictive mode, since the space between the plasma and the first wall is the vacuum, once we specify the plasma shape, the fluxes on the first wall can be calculated once for all (thus we avoid the need for ITER boundary code). Any full-domain equilibrium code could provide the necessary values of magnetic fluxes for given plasma shape on the first wall. (Fig. 4.)

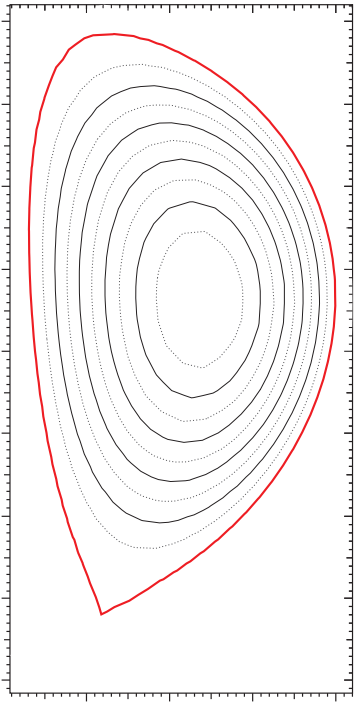

Fig 3.
ITER predictive equilibria: **the boundary of calculation domain** is determined and fixed by desired plasma shape.

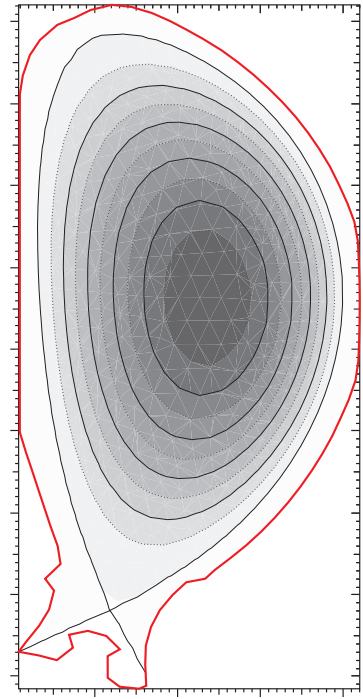

Fig. 4.
ITER first-wall based predictive equilibria: **the boundary of calculation domain** is first wall, values of the magnetic fluxes are calculated once for all by a full domain equilibrium code.

PROPOSED ADDITIONAL STABILIZATION OF THE PRESSURE PROFILE

Since the pressure component P of the source term is significantly more sensitive to the external and internal measurements than diamagnetic FF' term, we are planning the stabilisation of the pressure profile by introducing the values of the pressure on magnetic axis (PAX). The agreement of electronic density ( $n_e$ ) found by Equinox-J with direct measurement - the Thomson Scattering System (LIDAR) is already excellent (Fig. 5.):

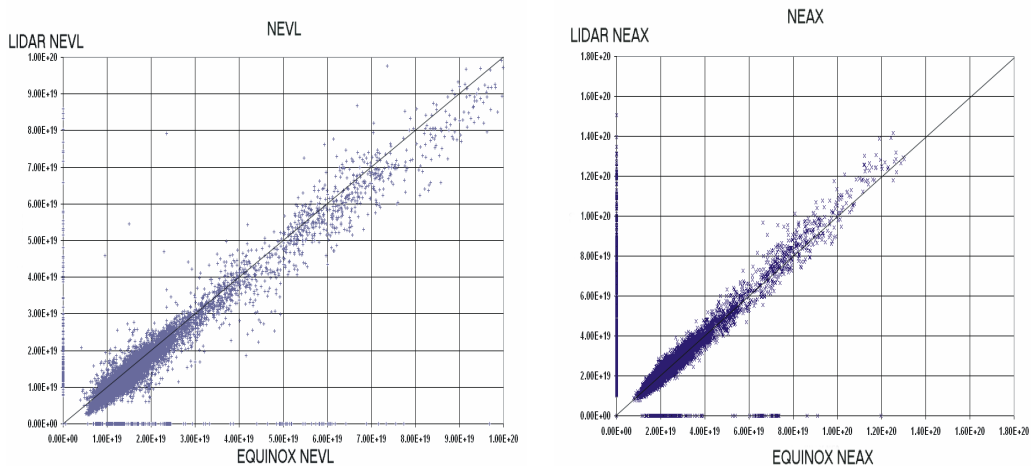

Fig. 5.

However the real-time measurement of pressure on axis is not generally available. Hopefully at JET we used successfully the formula:

$$P=(n_i T_i + n_e T_e) \qquad (7)$$

Where $n_i$ and $n_e$ are ionic and electronic densities

$T_i$ and $T_e$ are ionic and electronic temperatures

In order to estimate the pressure on axis we use offline direct Charge Exchange and LIDAR measurements together with electronic density from Equinox. A satisfactory fit to PAX obtained offline this time using real-time measurements has been obtained (Fig. 6).

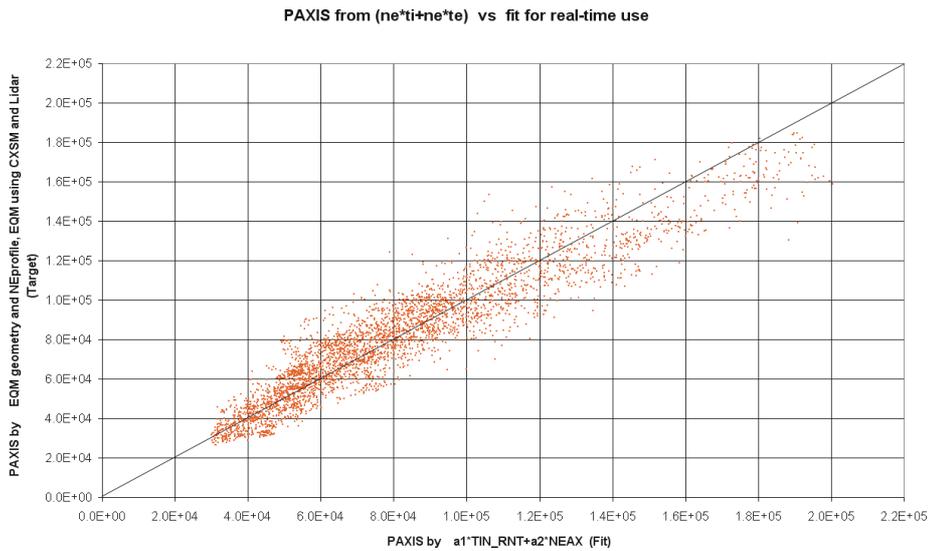

Fig. 6.

     The fit uses Total Neutron Rate and Electronic Density on Axis from real-time Equinox. The obtained value will be introduced directly in the equilibrium code in order to additionally constrain the pressure derivative. In practice this will require scaling of the pressure profile inside the equilibrium code.
     Unfortunately the real-time measurement is not available all the time during the pulse (mostly at high plasma current during 50% of pulse time and can be intermittent with 3-4 interruptions, 1s each). Rapid offset of poloïdal beta due to the switching the PAX signal on and off could be destabilizing, therefore some sort of smoothing on ramp-up and failsafe near-future prediction algorithm has to be developed to eliminate the effect of shaking the equilibrium.
     Another point is that this system would be highly destabilizing in case of overcorrected or undetected fringe jumps because of dependency on real-time electronic density – which is for now very robust with Equinox but may require improvements at the lower level at JET. For this reason it would be wise to include the more aggressive PAX stabilisation for Equinox-J (which requires very good electronic density anyway because it uses polarimetric measurements for fitting) and more conservative stabilisation to Equinox-M for better agreement with independent measurements (the latter is always converging and shows PAX already +/-20% at worst case compared to the PAX from our estimation).

CONCLUSION

- A careful implementation together with a new algorithm allowed a giant leap in execution time, towards a 6-12ms on an Anthlon 1700+ PC machine per time frame.
- EQUINOX code was tested on 5 operating systems, 3 different processor families (x86, Alpha, PXA250), 10 different compilers, >1000 different shots and installed for two tokamaks (JET, ToreSupra) with different geometry demonstrating its portability. Thanks to those tests infinitesimal numerical differences in the generated results were observed and the corresponding numerical methods have been tuned to increase the robustness.
- The code is mature and very robust, with particularly good worst-case results and completely unmanned operation.

- The future development will include normalized interface for several client-codes, f. ex. Matlab, Scilab, IDL and Maple codes, one of the first clients being CRONOS transport code, with both Linux and Windows operating system in mind.